\documentclass[notitlepage,superscriptaddress,aps,amsmath,amssymb,floatfix,reprint,twocolumn]{revtex4-1}
\usepackage{graphicx}
\usepackage[colorlinks=true,citecolor=blue,linkcolor=blue,urlcolor=blue]{hyperref}
\usepackage{amsmath}
\usepackage{dcolumn}
\usepackage{xcolor}
\usepackage{fancyhdr}
\usepackage{lipsum}
\usepackage{soul}

\usepackage{footnote}

\makeatletter 
\renewcommand{\fnum@figure}{{FIG.~\thefigure}}
\makeatother

\makeatletter
\def\bbordermatrix#1{\begingroup \m@th
  \@tempdima 4.75\p@
  \setbox\z@\vbox{%
    \def\cr{\crcr\noalign{\kern2\p@\global\let\cr\endline}}%
    \ialign{$##$\hfil\kern2\p@\kern\@tempdima&\thinspace\hfil$##$\hfil
      &&\quad\hfil$##$\hfil\crcr
      \omit\strut\hfil\crcr\noalign{\kern-\baselineskip}%
      #1\crcr\omit\strut\cr}}%
  \setbox\tw@\vbox{\unvcopy\z@\global\setbox\@ne\lastbox}%
  \setbox\tw@\hbox{\unhbox\@ne\unskip\global\setbox\@ne\lastbox}%
  \setbox\tw@\hbox{$\kern\wd\@ne\kern-\@tempdima\left[\kern-\wd\@ne
    \global\setbox\@ne\vbox{\box\@ne\kern2\p@}%
    \vcenter{\kern-\ht\@ne\unvbox\z@\kern-\baselineskip}\,\right]$}%
  \null\;\vbox{\kern\ht\@ne\box\tw@}\endgroup}
\makeatother
\begin{document}

\title{Magneto Acoustic Spin Hall Oscillators}
\author{Mustafa  Mert Torunbalci}
\email{mtorunba@purdue.edu}
\affiliation{Purdue University, School of Electrical and Computer Engineering, West Lafayette, IN, 47907, USA}
\author{Tanay A. Gosavi}
\affiliation{Cornell University, School of Electrical and Computer Engineering, Ithaca, NY, 14853, USA}
\author{Kerem Y. Camsari}
\affiliation{Purdue University, School of Electrical and Computer Engineering, West Lafayette, IN, 47907, USA}
\author{Sunil A. Bhave}
\affiliation{Purdue University, School of Electrical and Computer Engineering, West Lafayette, IN, 47907, USA}
\title  {Magneto Acoustic Spin Hall Oscillators}
\begin{abstract}This paper introduces a novel oscillator that combines the tunability of spin Hall-driven nano oscillators with the high quality factor (Q) of  high overtone bulk acoustic wave resonators (HBAR), integrating both reference and tunable oscillators on the same chip with CMOS. In such magneto acoustic spin Hall (MASH) oscillators, voltage oscillations across the magnetic tunnel junction (MTJ) that arise from a spin-orbit torque (SOT) are shaped by the transmission response of the HBAR  that acts as a multiple peak-bandpass filter and a delay element due to its large time constant, providing delayed feedback. The filtered voltage oscillations can be fed back to the MTJ via a) strain, b) current, or c) magnetic field. We develop a SPICE-based circuit model by combining experimentally benchmarked models including the stochastic Landau-Lifshitz-Gilbert (sLLG)  equation for magnetization dynamics and the Butterworth Van Dyke (BVD) circuit for the HBAR. Using the self-consistent model, we project up to $\sim$ 50X enhancement in the oscillator linewidth with Q reaching up to 52825 at 3 GHz, while preserving the tunability by locking the STNO to the nearest high Q peak of the HBAR.  \linebreak We expect that our results will inspire MEMS-based solutions to spintronic devices by combining attractive features of both fields for a variety of applications.
\end{abstract}

\maketitle

\section*{Introduction}
Frequency synthesizers are essential building blocks for  modern communication systems such as cell phones, radio receivers, TVs, and GPS.  These devices consist of a reference quartz crystal oscillator that is phase locked to a voltage controlled oscillator (VCO) through a frequency divider and are used to produce a range of stable output frequencies.  However, off-chip components used to implement these devices consume large power and area.  MEMS oscillators are promising candidates to replace quartz crystal oscillators in frequency synthesizers since they are CMOS-compatible, can have kHz to GHz operation, have low phase noise and excellent stability \cite {ruby2012positioning, nguyen2007mems, basu2011microelectromechanical, fedder2008technologies, qu2016cmos, li2013cmos, chen2011generalized, fischer2016integrating}. However, they suffer from extremely limited tunability, less than 100 ppm at GHz frequencies, severely restricting their use as a single chip oscillator in the communication systems \cite {van2011review}.  

Recent achievements in spintronics have enabled development of nanoscale, CMOS-compatible, GHz operation, and tunable spin torque nano oscillators (STNO) \cite { kiselev2003microwave, dumas2014recent, chen2015integration, chen2016spin, makarov2016cmos}.  These properties make STNOs promising candidates for communication applications.  However, STNOs suffer from low output power and large linewidth \cite {quinsat2010amplitude}.   Output power of the STNO can be improved by synchronizing a couple of STNOs or making some structural optimizations \cite {kaka2006mutual, slavin2009nonlinear, ruotolo2009phase, sani2013mutually, locatelli2015efficient, lebrun2017mutual, banuazizi2017order}.  Linewidth of the STNO can be reduced using various ways.   Injection locking is the most well-known technique where an external signal is injected to reduce the linewidth \cite { rippard2005injection, burgler2011injection, rippard2013time, gosavi2015model, ganguly2016mesh}.   An alternative approach is to use a self-delayed feedback in which the current of the STNO is reinjected after a certain delay, reducing the linewidth and critical current needed for oscillations \cite {tiberkevich2014sensitivity, khalsa2015critical, tsunegi2016self, dixit2012spintronic, kumar2016coherent, tamaru2015extremely, kreissig2017vortex, bhuktare2017spintronic}.  Alternatively, a multiple peak-high Q HBAR filter is used to reduce the linewidth of the STNOs in the open loop configuration \cite{gosavi2017magneto}.  These methods are intended to improve the linewidth of a free running STNO.  However, it is also necessary to develop a method that not only improves the linewidth but also provides the integration of the STNO and feedback components on the same chip for a single chip tunable oscillator.  

In this paper, we propose MASH oscillators that consist of a three terminal MTJ and an HBAR on the same silicon substrate with CMOS circuits, constituting a single chip oscillator. We have also shown that two-chip version of the MASH oscillators, where HBAR is implemented on a lower loss substrate, show further reduction in the linewidth at the cost of being CMOS-incompatible.  Figure 1 summarizes our vision of MASH oscillators with strain, current, or magnetic field feedback.  We first develop a model by solving the sLLG equation for magnetization dynamics and transport equations for spin Hall effect (SHE) and MTJ self-consistently with the BVD circuit for the HBAR in a unified SPICE based circuit platform.  Each individual component of the model (sLLG, SHE, MTJ, and BVD) is experimentally benchmarked or equivalent to the-state-of the art theoretical prescription \cite{camsari2015modular}.  Using this model, we compare the proposed MASH oscillators with a free running STNO using identical STNO and HBAR modules. MASH oscillators exhibit up to 50X enhancement in the linewidth, while maintaining the tunability by locking the STNO to the nearest high Q peak of the HBAR.

\begin{figure*}[ht]
\centering
\includegraphics [width=0.95\linewidth]{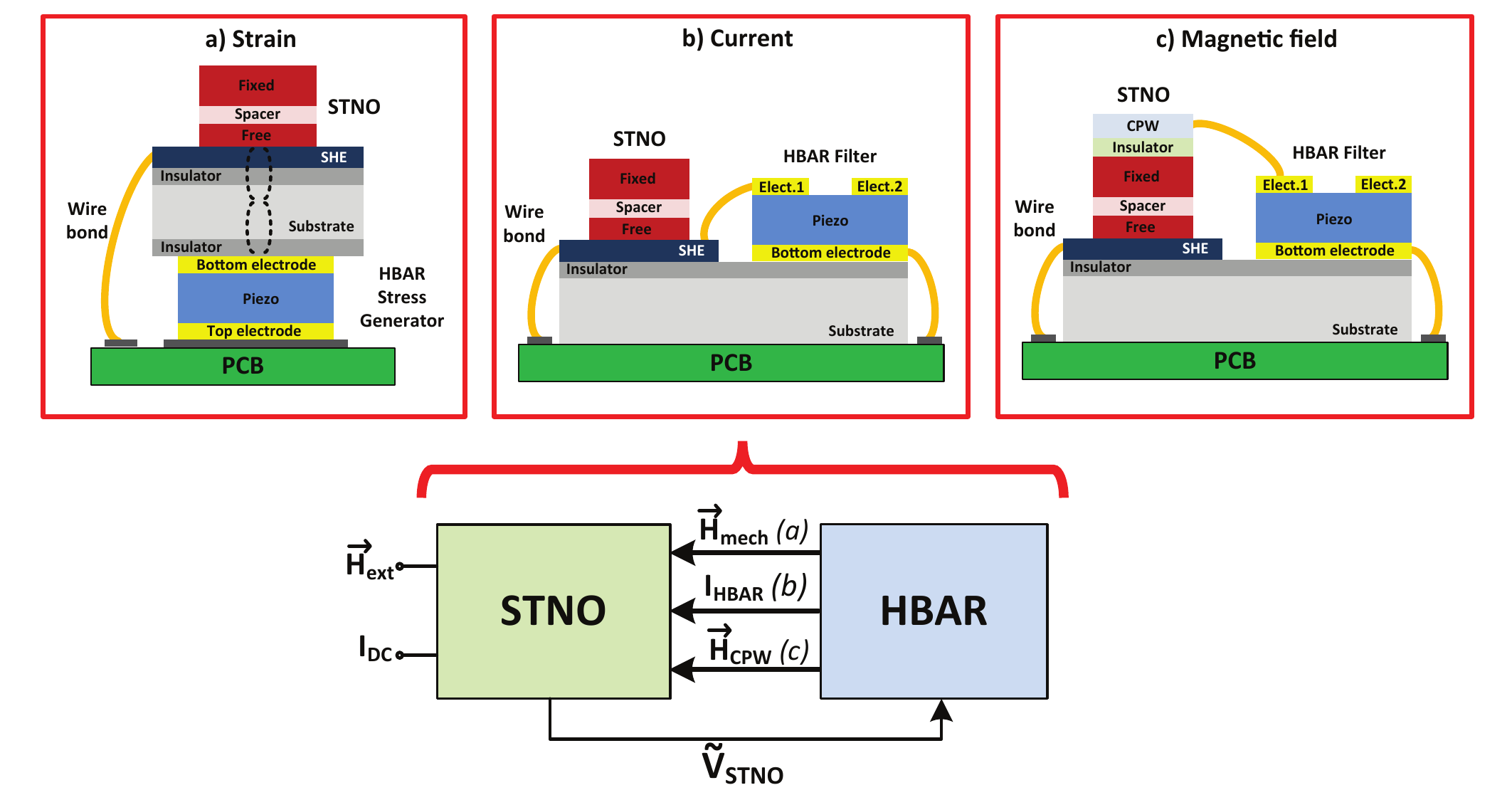}
\caption{\textbf{Magneto acoustic spin Hall oscillators} that combine the tunability of the standard STNO with high Q of the HBAR on a single chip. Voltage oscillations across a three-terminal MTJ is shaped by the transmission response of the HBAR and fed back to the MTJ after a certain delay using a) strain, b) current, or c) magnetic field via a CPW (coplanar waveguide), causing high Q filtering and delayed feedback.  Proposed concept shows a significant enhancement in the linewidth with all feedback methods, simultaneously preserving the key frequency tunability feature of STNOs.}
\label{fig:Figure1}
\end{figure*}

\section*{Simulation Framework}
In this section, we describe our SPICE-based simulation framework consisting of sLLG and BVD modules used to implement the proposed MASH oscillator model.  Identical STNO and HBAR modules are used to compare MASH oscillators with strain, current, and magnetic field feedback.  Strain feedback requires an STNO and a 1-port HBAR stress generator on the opposite sides of the substrate whereas current and magnetic field feedback uses an STNO and 2-port HBAR bandpass filter on the same side.  
\subsubsection*{Description of STNO and HBAR modules}
 \paragraph{STNO module.} STNO consists of two ferromagnetic layers, a fixed and a free layer, separated by an insulator spacer.  The free layer of the SNTO is driven into precessional dynamics by an external magnetic field and a spin polarized current that are sensed by an MTJ as voltage oscillations. We model STNOs by a standard LLG equation in the monodomain approximation using the spin-circuit framework developed in Ref. \cite {camsari2015modular}.  The spin-circuit approach allows us to combine the SHE, MTJ, and MEMS circuits in a unified SPICE based circuit platform.  Figure 2 (a) presents  modular modeling of transport and magnetization dynamics for the STNOs.  The time response of the magnetization along the easy axis is calculated using the sLLG equation:   
\begin{eqnarray}
 && \footnotesize (1+\alpha^2)\frac{d\hat{m}_i}{dt}=-|\gamma|\hat{m}_i\times \vec{H}_{eff} -\alpha|\gamma|( \hat{m}_i \times \hat{m}_i \times \vec{H}_{eff}) \nonumber \\ 
&& \footnotesize +\frac{1}{qN_s}(\hat{m}_i \times\vec{I}_{S}\times\hat{m}_i  )+\frac{\alpha}{qN_s}(\hat{m}_i \times\vec{I}_{S})
\end {eqnarray}
where $\hat{m}_i$ is the unit vector along the magnetization, $\gamma$ is the gyromagnetic ratio, $\alpha$ is the damping constant,  $\vec{I}_{S}$ is the total spin current, defined as the vectorial sum of SHE ($\Vec{I}_{SHE}$) and MTJ ($\Vec{I}_{stt}$) components. $N_s$ is the total number of spins given by $N_s$=$M_s$V/$\mu_B$ where $M_s$ is the saturation magnetization, $V$ is the volume of the free layer, $\mu_B$ is the Bohr magneton, and  $\vec{H}_{eff}$ is the effective magnetic field including the uniaxial, shape anisotropy and magnetic thermal noise terms.  The thermal noise ($\vec{H}_{n}$) enters as an uncorrelated external magnetic field in three dimensions with the following mean and variance, equal in all three directions \cite{sun2004spin}: 
\begin{equation}
\big \langle {H}^{\Vec{r}}_{n} \big \rangle=0 \quad \textrm{and} \quad   \big \langle |{H}^{\Vec{r}}_{n}|^2 \big \rangle= \displaystyle\frac{2\alpha kT}{\gamma M_s V}
\end {equation}
where $k$ is the Boltzmann constant and $T$ is the temperature.  In order to achieve independent control of oscillation amplitudes and frequency \cite{liu2012magnetic}, we assume that the spin current is generated by SHE where charge to spin current conversion ($\beta$) is expressed by \cite{hong2016spin}: 
\begin{equation}
\beta=\frac{I_{SHE}}{I_{c}}=\theta_{SHE}\frac{l_{SHE}}{t_{SHE}}\bigg(1-\sec\bigg[\frac{t_{SHE}}{ \lambda_{sf}}\bigg]\bigg)
\end {equation}
where $\theta_{SHE}$ is the spin Hall angle, $l_{SHE}$ and $t_{SHE}$ are the length and thickness, and $\lambda_{sf}$ is the spin-flip length of the SHE metal.  We assume that the SHE spin-current is absorbed with 100\% efficiency for simplicity and its spin polarization is along $\hat{z}$ since charge flow is on $\hat{y}$ and surface normal is along $\hat{x}$.  Recent developments in the SHE have enabled the use of materials with large $\theta_{SHE}$ such as tungsten corresponding an efficient charge to spin current conversion \cite{pai2012spin, hao2015giant}.  Therefore, the conversion efficiency can exceed 1, providing an intrinsic gain. For the parameters used throughout the paper, $\beta$ is 2.1.   We use a simple bias independent three-terminal MTJ model for CoFeB (Fixed) /MgO/CoFeB (Free)/W stack where both free and fixed CoFeB layers are magnetized in-plane in the $\hat{z}$ direction, assuming a TMR of 112\% with $R_p$=400 $\Omega$ and $R_{ap}$=850 $\Omega$. RA product of the MTJ is assumed to be 4 $\Omega \mu m^2$.  Assuming an external field in the $\hat{z}$ direction, the effective magnetic field becomes $\vec{H}_{eff}$=($\vec{H}_{an}$$m_z$+ $\vec{H}_{ext}$)$\hat{z}$$-$ $\vec{H}_{d}$$ m_x\hat{x}$ +$\vec{H}_{n}$.  We have also assumed that the stray field due to the fixed layer is subsumed into the externally applied field, $\vec{H}_{ext}$ since these are along the same direction.  Voltage oscillations across the MTJ can be expressed as:
\begin{equation}
V_{STNO}=V_{read} \frac{R_{MTJ}}{R_{MTJ}+R_{read}}
\end {equation}
where $R_{MTJ}=1/[G_0(1+P^2 m_z)]$ \cite {camsari2014physics}, $G_0$ is average MTJ conductance ([$G_{P}$+$G_{AP}$]/2), $P$ is polarization, $V_{read}$, and $R_{read}$ the read voltage and resistance. The fixed layer contribution to the spin-current is:  $\Vec{I}_{stt}=PI_{read}$. The resistance contribution of the SHE  is neglected in the equivalent read circuit since the MTJ resistance is the dominant resistance in this path.  Figure 2 (b) shows simulated FFT (Fast Fourier Transform) spectrums of a free running STNO at  T=300 K and T=0 K (noiseless)  for $V_{tune}$=23 mV, showing the fundamental mode and harmonics.   Figure 2 (c) shows the tunability of the STNO that can be independently controlled by $\vec{H}_{ext}$ and $V_{tune}$.  The oscillation frequency increases with the square root of the external magnetic field at low fields and linearly at high fields as predicted by the Kittel equation $f_0=\gamma / 2\pi \sqrt{(H_{ext}+H_{an})(H_{ext}+H_{an}+H_{d})}$.  Alternatively, the oscillation frequency can be tuned by a charge current passing through SHE metal generating a spin current on the free layer. There are two main modes of vibration for the STNOs: in-plane and out-of plane modes. In the in-plane mode, frequency decreases as $V_{tune}$ increases whereas the opposite is valid for the out-of-plane mode \cite { kiselev2003microwave}. In order to sustain periodic oscillations, the spin-torque needs to exceed a critical value to overcome the damping terms, which in this work corresponds to a $V_{tune}$ of 15 mV.  There are no oscillations below this value. STNO operation requires the spin-torque current and external field to be in anti-parallel directions to sustain continuous oscillations. Therefore simply changing the polarity of the tuning voltage to negative values do not result in oscillations (and hence not shown), unless the magnetic field direction is also reversed. 

\begin{figure*}[ht]
\centering
\includegraphics[width=\linewidth]{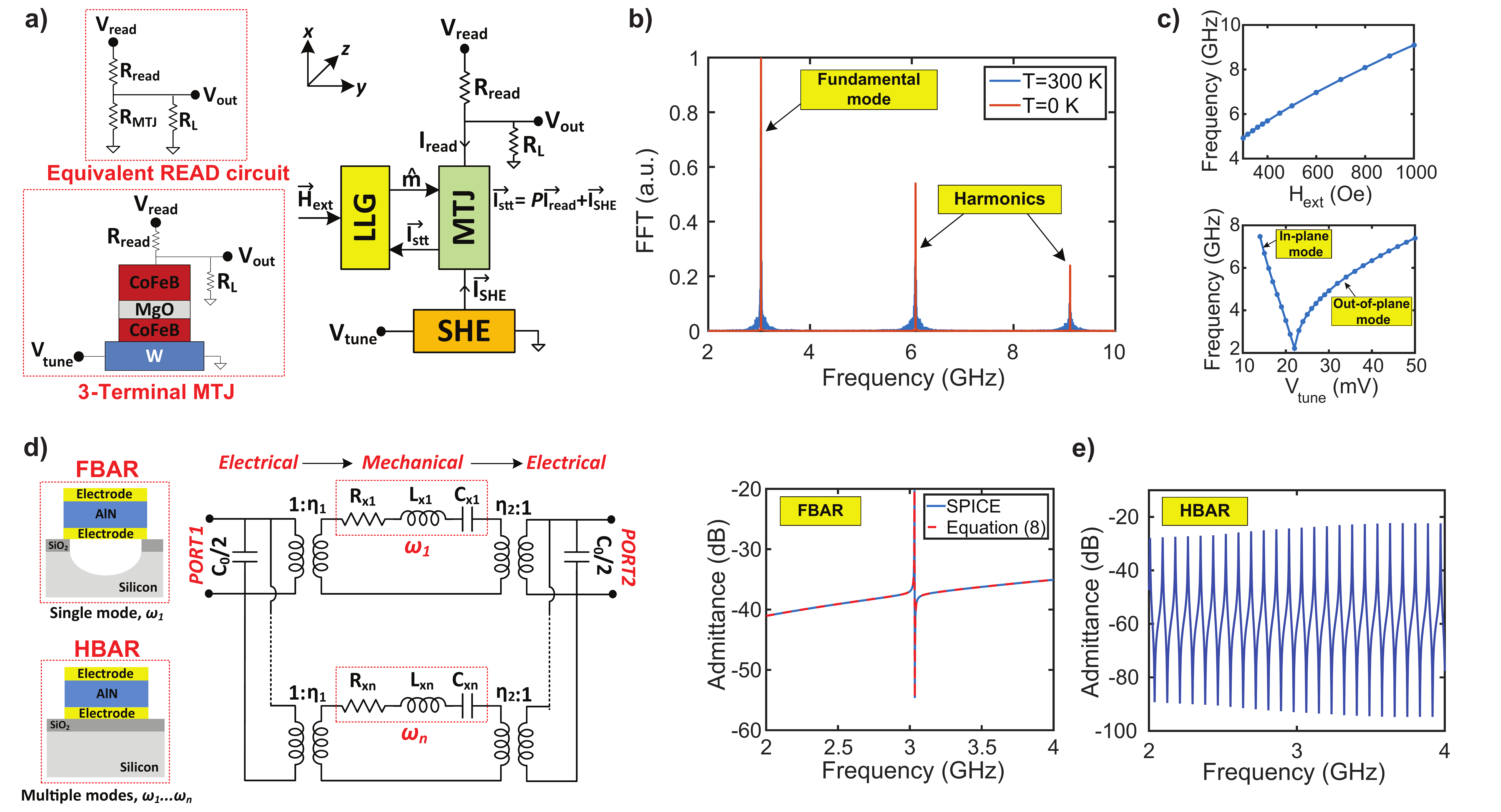}
\caption {\textbf{Decoupled STNO and HBAR module:} a) Modular modeling of transport and magnetization dynamics of the STNO. b) Simulated FFT spectrums of a free running STNO at T=300 K and T=0 K (noiseless) for $V_{tune}$=23 mV, showing the fundamental mode and harmonics.  c) Tunability of a free running STNO with $\vec{H}_{ext}$ and $V_{tune}$. d) Equivalent circuit model for the HBAR where each acoustic mode is modeled as a separate RLC circuit. e) Admittance plots of a single mode resonator (FBAR) and a 2-port HBAR filter, consistent with the Equation (8). HBAR has multiple high Q resonance peaks that are spaced by $\Delta f$=43 MHz depending on the substrate thickness.}
\label{fig:MASH}
\end{figure*}

\paragraph{HBAR module.}HBAR is a key element in MASH oscillators since it provides high Q resonance peaks, enabling high Q filtering with an effective delay and maintaining tunability of the STNO.  The HBAR consists of a piezoelectric layer sandwiched between two electrodes on an acoustic substrate.  The piezoelectric film acts as a transducer to generate standing waves at several wavelengths through the acoustic substrate.  The acoustic substrate acts as a resonant cavity so that several resonance peaks occur that are separated from each other by a known frequency spacing, $\Delta f$.  We use different HBAR designs with a 1 $\mu$m AlN layer on top of a 100 $\mu$m and 200 $\mu$m thick silicon substrates: a) 1-port HBAR stress generator for the strain feedback and b) 2-port HBAR filter to implement the current and magnetic feedback.  The frequency of the HBAR is $f_{rn}\approx[(n+1)/2t_{Si}] \sqrt{E_{Si}/\rho_{Si}} $ where n is number of the acoustic modes, $t_{Si}$ is silicon thickness, $E_{Si}$ and $\rho_{Si}$ are the Young's modulus and density of the silicon.   We neglect the contribution of AlN layer on the equivalent mass and spring constant in the model since HBAR is mostly formed by the silicon substrate.  This assumption is valid as long as the thicknesses of the piezoelectric film is much smaller than the thickness of the substrate.  The fundamental mode of the HBAR is when a full wavelength fits to the AlN layer, $t_{AlN}=\lambda /2$. Total number of the modes can be found by $n= t_{cavity}/(\lambda/2)$ where $t_{cavity}=t_{AlN}+t_{Si}$, which is equal to 101 modes that are separated by $\Delta f$=$(1/2t_{Si}) \sqrt{E_{Si}/\rho_{Si}}$=43 MHz for a 100 $\mu$m thick HBAR design.  Figure 2 (d) presents the BVD circuit model of the HBAR where each acoustic mode is defined as a separate RLC circuit \cite{tilmans1996equivalent}.  The mechanical equivalent of $R_{xn}$, $L_{x}$, and $C_{xn}$ is the damping constant ($b_{xn}$),  equivalent mass ($m_x$), and inverse of equivalent spring constant (1/$k_{xn}$), respectively.  The $R_{xn}$ is also equal to  $\sqrt{k_{xn}m_{x}}/Q_{n}$ where $Q_{n}$ is the mechanical quality factor of a defined mode.  We have experimentally shown that  the $fQ$ product of a silicon HBAR is approximately $10^{13}$ in \cite {gosavi2015hbar} which provides Q of 2500-5000 within the frequency range of 2-4 GHz for the model.   The $ k_{xn}$ and  $m_{x}$ can be calculated by:
\begin{equation}
k_{xn}={(n+1)}^2 \frac{E_{Si} \pi ^2 A } {2t_{Si}}\quad m_{x}=\frac{1}{2} \rho_{Si} A t_{Si}
\end {equation}
The electrical to mechanical conversion in the BVD circuit is expressed by two transformers where $\eta_1$ and $\eta_2$ are the transduction factors, which are typically equal to each other ($\eta_{HBAR}$=$\eta_1$=$\eta_2=2e_{33} A/t_{AlN}$) for symmetric structures where where $e_{33}$ is the out-of plane piezoelectric coefficient of AlN, $A$ is the area of the HBAR, $t_{AlN}$ is the thickness of the AlN film \cite{tilmans1996equivalent}. The capacitance between the top and bottom electrodes of the HBAR is equal to $C_0=\epsilon_{0}\epsilon_{AlN}A/t_{AlN}$ where $\epsilon_{0}$ and $\epsilon_{AlN}$ are the dielectric constant of the free space and AlN layer.

The BVD circuit can be used to model 1-port stress HBAR generator by removing the transformer at the output.  Therefore, the output current of the HBAR becomes equal to the mechanical velocity $(\partial x/\partial t)$ and can be converted to the mechanical displacement ($x$) with an integrator.  Assuming the stress is uniform through the substrate, strain is approximately equal to $S$=$x/t_{si}$.  The 2-port HBAR is modeled using the BVD circuit with two output electrodes. The 2-port configuration can be explained by two coupled HBAR devices where transmission loss is determined by the gap between the electrodes. We use identical parameters for both 1-port and 2-port HBAR designs in order to make an accurate comparison between different feedback methods. Figure 2 (e) compares admittance plots of a film bulk acoustic wave resonator (FBAR) and a 2-port HBAR, consistent with the Equation (8) and our experimental results in \cite{gosavi2015hbar, gosavi2017magneto}.  Unlike an FBAR, HBAR has multiple high Q resonance peaks that are spaced by $\Delta f$=43 MHz for a 100 $\mu$m thick silicon substrate.  The HBAR has a large time constant and acts as a delay element where the effective delay can be calculated as $\Delta t=(2Q_{n})/w_{rn}$, which is 330 ns at 3 GHz with a Q of 3000.  
\section*{Results}
\subsubsection*{MASH oscillator with strain feedback}
 Figure 3 (a-b) shows the block diagram and modular modeling of MASH oscillator implemented with an STNO and a 1-port HBAR stress generator on the opposite sides of the silicon substrate with strain feedback.  Magnetization of the free layer is controlled using an AC stress generated by the HBAR with magnetostriction effect \cite{ gowtham2016thickness, gowtham2015critical, khan2014voltage, roy2012energy}.   In the presence of a uniaxial mechanical stress and external magnetic field in the $\hat{x}$ and $\hat{z}$ directions, effective magnetic field  becomes $\vec{H}_{eff}$=($\vec{H}_{an}$$m_z$ + $\vec{H}_{ext}$)$\hat{z} $$-$$ (\vec{H}_{d}+\vec{H}_{mech})$$m_x$$ \hat{x}$ +$\vec{H}_{n}$.  The periodic oscillations on the free layer is sensed by the MTJ as an oscillation voltage.  This voltage $(V_{STNO})$ is amplified by a gain stage to bias the HBAR in order to generate sufficient amount of stress in the $\hat{x}$ direction.  Output current of the HBAR driven by the STNO can be written as:
\begin{equation}
\scriptsize I_{HBAR}=A_{amp}V_{STNO}\bigg(jwC_{0}+ \frac {jwC_x\eta_{HBAR}}{({jw_{rn})}^2L_{x}C_{x}+1+jwR_{x}C_{x}}\bigg)
\end {equation}
When driven at resonance ($w$=$w_{rn}$=1/$\sqrt{L_{xn}C_{xn}}$),  HBAR becomes a low impedance path with $Z$($w_{rn}$)=$R_{xn}$/$\eta_{HBAR}$.  Therefore, the output current becomes $I_{HBAR}$=($A_{amp}$$V_{STNO}$$\eta_{HBAR}$)/$R_{xn}$ and equal to the mechanical velocity ($I_{HBAR}$=$(\partial x/\partial t)$). The mechanical velocity can be converted to the mechanical displacement ($x$) with a simple integration. Assuming that the stress is uniform through the substrate, the strain ($S$) approximately is equal to $S$=$x/t_{si}$. The strain injects an AC magnetic field on the free layer via \cite{gowtham2015traveling}: $\vec{H}_{mech}=2B_{eff}S/M_{s}$  where $B_{eff}$  is the magnetoelastic coupling coefficient.  The magnetoelastic coupling coefficient for 2 nm CoFeB is measured to be $-9\times$$10^{7}$ erg/$cm^3$  in \cite{gowtham2016thickness}, and the HBAR can generate AC strain in the range of 20-200 ppm depending on the output voltage of the SNTO and amplifier gain. This amount of strain injects an AC magnetic field of 4.5-45 $Oe$ on the free layer.   
\begin{figure*} 
\centering
\includegraphics   [width=\linewidth] {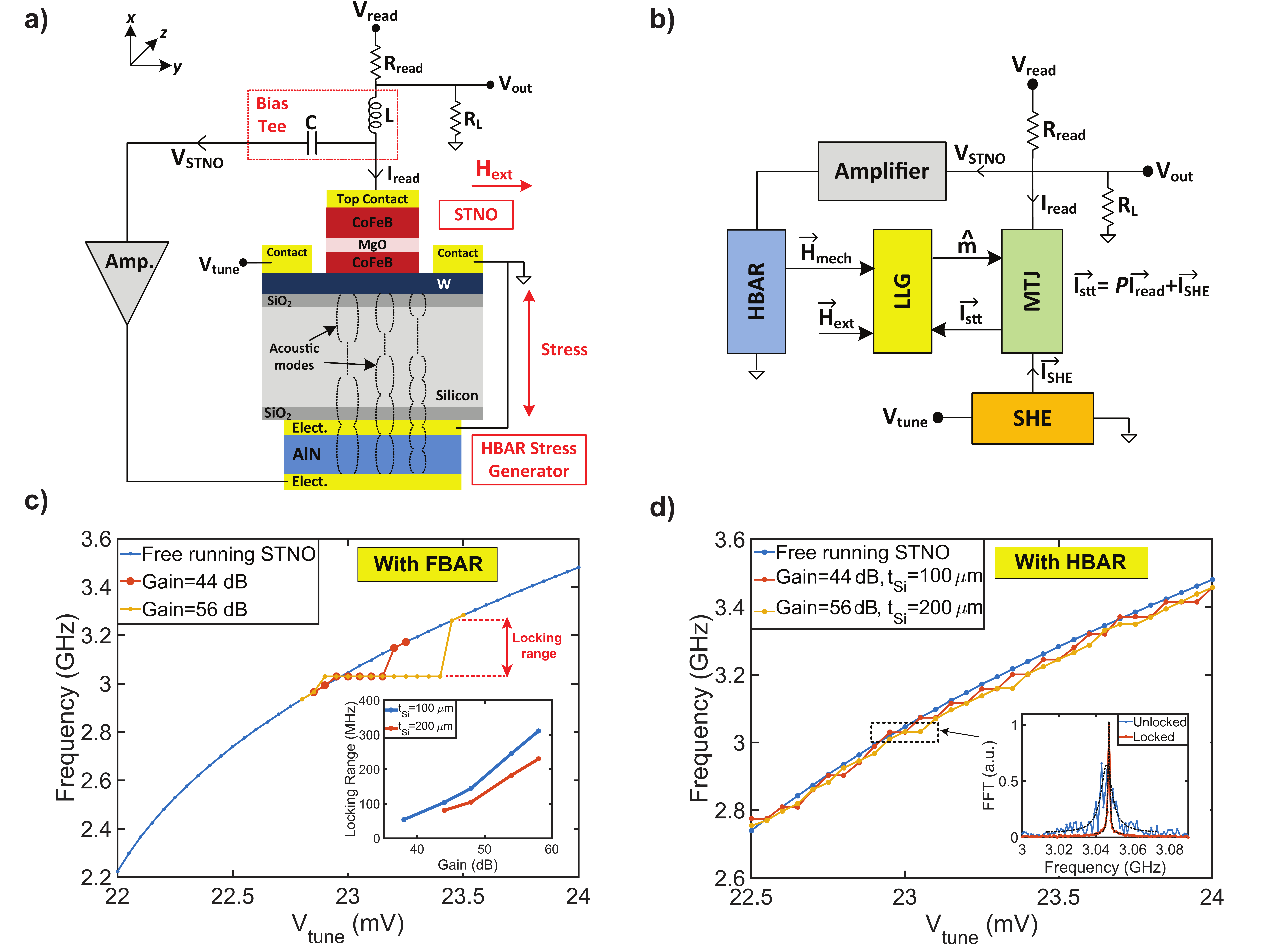}  
\caption{\textbf{Implementation of MASH oscillator with strain feedback:} a) A three-terminal MTJ and 2-port  HBAR are fabricated on the opposite sides of the same silicon substrate, and coupled to each other with strain due to the magnetostriction effect.  (b)  Modular modeling of transport, magnetization, and resonator dynamics. c) Locking range versus amplifier gain for two different substrate thicknesses (100 $\mu$m and 200 $\mu$m) simulated with an FBAR where thicker substrate ($t_{Si}$= 200 $\mu$m) requires larger gain since it can generate lower strain. (d) Tunability is achieved with $V_{tune}$ by locking the STNO to the nearest HBAR peak.  The thickness of the substrate determines the tuning-steps by defining the spacing between the HBAR modes.  }
\label{fig:strain}
\end{figure*}
The injected magnetic field causes a phase shift in the STNO and the oscillation frequency changes to compensate the phase shift, causing the locking.  When locked, the phase of the STNO follows the phase of the injection signal \cite {adler1946study, razavi2004study}.  Since the HBAR continuously cleans the STNO signal and generates a high Q injection signal in the loop, the linewidth of the STNO improves with the simultaneous effect of high Q filtering and delayed feedback by the HBAR.  Figure 3(c) presents the locking range of the STNO for different amplifier gains and substrate thicknesses (100 $\mu$m and 200 $\mu$m) simulated with an FBAR.  The locking range can be calculated by using Adler's equation: 
\begin{equation}
\Delta w=\frac{w_{0}}{2Q} \frac{I_{inj}}{I_{osc}}  \frac{1}{\sqrt{1-\displaystyle\frac{I_{inj}^2}{I_{osc}^2}}}
\end {equation}
where $\Delta w$ is the maximum locking range,  $w_0$ is oscillation frequency, $Q$ is quality factor, $I_{osc}$ is oscillation current, and $I_{inj}$ is injected signal for the locking.  Equation (7) can be simplified to $\Delta w$=($w_{0}$/2$Q$)($I_{inj}$/$I_{osc}$) if $I_{inj}$ $\ll$ $I_{osc}$, predicting a linear locking range with  increased $I_{inj}$.  However, MASH oscillator produces its own $I_{inj}$ by driving the HBAR at or near resonance.  Therefore, $I_{inj}$ does not linearly increase with increased amplifier gain, showing a non-linear locking range. Amplifier gain can be reduced by increasing the output voltage of the STNO or using a material stack with higher magnetostrictive and piezoelectric coefficients.  The substrate thickness ($t_{Si}$) also affects the gain needed for locking since the thicker substrate ($t_{Si}$=200 $\mu$m) can generate lower strain.

Figure 3 (d) presents tunability of MASH oscillator with the strain feedback where the STNO is locked to the nearest peak of the HBAR that are separated by 43 MHz for a 100 $\mu$m thick HBAR.  Therefore, MASH oscillator can be tuned with 43 MHz steps by changing $V_{tune}$. When locked, MASH oscillator demonstrates a significant enhancement in the linewidth with the combination of high Q filtering and delayed feedback. The tuning step can be decreased if the spacing between the HBAR modes are reduced.  For instance, separation between the modes becomes 22 MHz  when $t_{Si}$ is 200 $\mu$m, showing a step tunability of 22 MHz.  However, this also decreases the amount of stress generated by the HBAR, requiring higher amplifier gain for the feedback as shown in the Figure 3 (d).

\subsubsection*{MASH oscillator with current feedback}
 Figure 4 (a-b) presents the implementation of MASH oscillator with a three-terminal MTJ and a 2-port HBAR using the current feedback.   Voltage across the MTJ $(V_{STNO})$ is amplified by a gain stage and filtered by a 2-port HBAR filter, allowing the signal transmission only at its sharp resonance peaks.  The filtered current output of the HBAR can be expressed as:   
\begin{equation}
\footnotesize I_{HBAR}=A_{amp}V_{STNO} \frac{1}{k_{xn}} \frac {{jw\eta_{HBAR}}^2}{\left (\displaystyle j\frac{w}{w_{rn}}\right)^2+ 1 + \displaystyle j\frac{w}{w_{rn} Q_{n}}}
\end {equation}
where $A_{amp}$ is amplifier gain, $w_{rn}$ is the frequency of the HBAR, and $V_{STNO}$ is the output voltage of the STNO. When driven at resonance ($w$=$w_{rn}$=$\sqrt{k_{xn}/m_{x}}$), HBAR becomes a low impedance path with $Z$($w_{rn}$)=$R_{xn}$/$\eta_{HBAR}^2$.  Therefore, the injected AC current is  $I_{inj}$=($A_{amp}$$V_{STNO}$ $\eta_{HBAR}^2$)/($R_{xn}$+$R_{SHE}$) that is used to generate an AC spin current through the SHE metal ($I_{sinj}$=$I_{inj}$$\beta$), causing an injection locking.  Even though the geometric factor ($l_{SHE}$/$t_{SHE}$) in Equation (3) can be used to provide an intrinsic gain in the loop, this gain is not sufficient.  Therefore, an additional amplifier is used to compensate the transmission losses and provide the gain necessary for the feedback.  

\begin{figure*}
\centering
\includegraphics[width=0.825\linewidth] {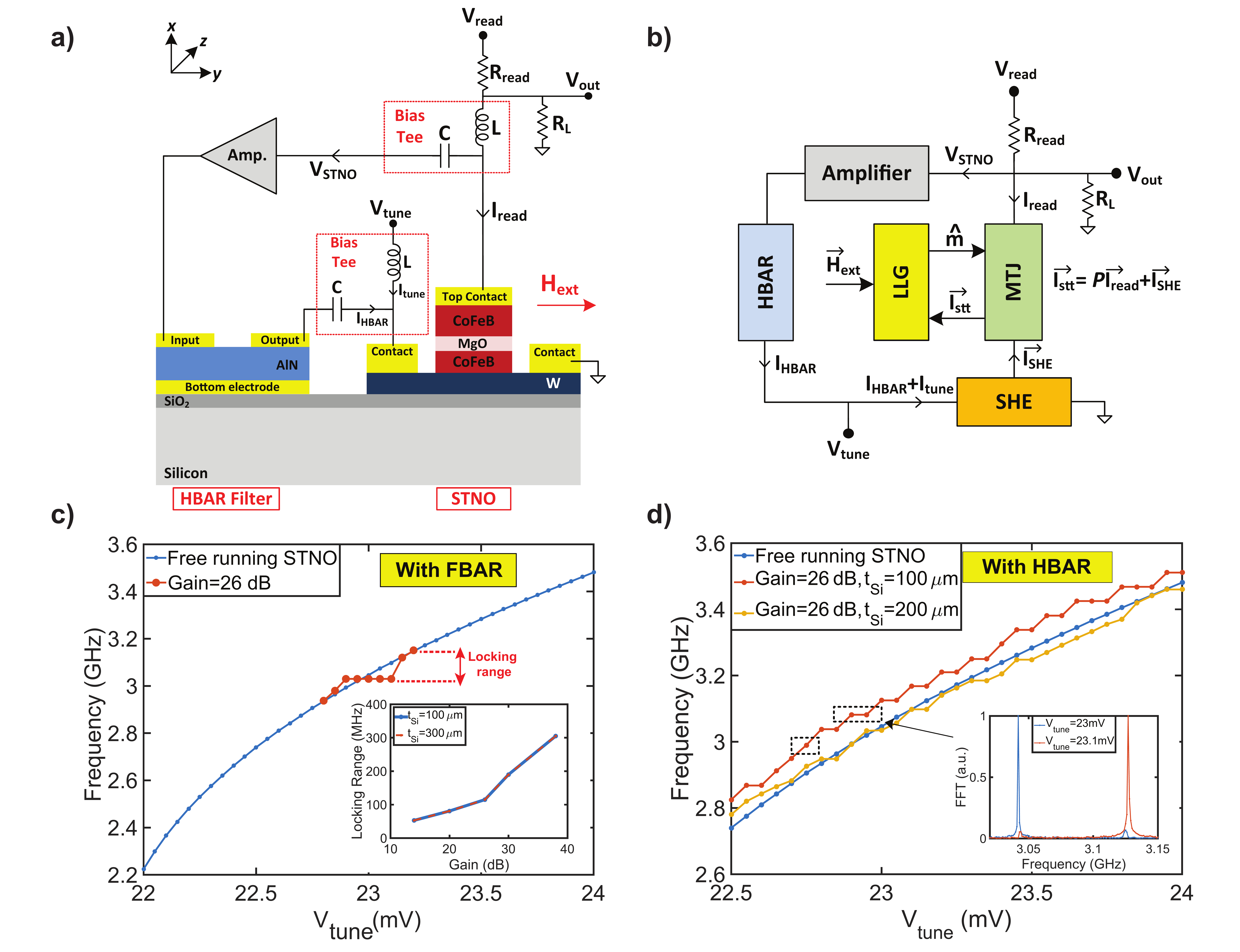}
\caption{\textbf{Implementation of MASH oscillator with current feedback:} a) A three-terminal MTJ and a 2-port  HBAR are fabricated on the same silicon substrate, and coupled to each other with the current feedback.   (b)  Modular modeling of transport, magnetization, and resonator dynamics. c) Locking range versus amplifier gain for two different substrate thicknesses (100 $\mu$m and 200$\mu$m) simulated with an FBAR. The thicker substrate ($t_{Si}$=200 $\mu$m) does not affect the gain needed for the locking. (d) Tunability can be achieved with 22 MHz or 43 MHz steps depending on the substrate thickness ($t_{Si}$) using the same amplifier gain by locking the nearest high Q peaks of the HBAR.}
\label{fig:Current feedback}
\end{figure*}
Figure 4 (c) presents  the locking range of the STNO for different amplifier gains and substrate thicknesses (100 $\mu$m and 200 $\mu$m) simulated with an FBAR.  Similar to strain feedback, we do not have a linear locking range since $I_{inj}$ shows non-linear characteristics by driving the HBAR at or near resonance.  Nevertheless, the simulation results seem consistent with Equation (7) and experimental results where injection locking is achieved by an external current source \cite{rippard2005injection, rippard2013time}. Compared to the strain feedback, a lower amplifier gain is sufficient to reach a similar locking range and it does not depend on the substrate thickness ($t_{si}$), proving that the current feedback is more efficient.  Amplifier gain can further be decreased by reducing the transmission loss that depends on the lateral spacing between the input and output electrodes of the HBAR.  Figure 4 (d) shows the tunability of MASH oscillator where the output frequency can be tuned with 43 MHz steps by changing $V_{tune}$ for a 100 $\mu$m thick HBAR similar to the strain feedback.  Using a thicker substrate ($t_{Si}$=200 $\mu$m) reduces the separation ($\Delta f$=22 MHz) between the HBAR modes, and this enables a tunability with 22 MHz steps without increasing the amplifier gain.

\subsubsection*{MASH oscillator with magnetic field feedback}
Figure 5 (a-b) shows the implementation and circuit model of MASH oscillator with an STNO including a CPW and a 2-port HBAR filter using the magnetic field feedback.  Voltage oscillations across the MTJ is shaped by the HBAR.  Output current of the HBAR is passed through the CPW after a certain delay to injects an AC magnetic field on the free layer, providing the magnetic field feedback.  In the presence of the magnetic feedback, effective field becomes $\vec{H}_{eff}$=($\vec{H}_{an}$$m_z$ + $\vec{H}_{ext}$)$ \hat{z} - (\vec{H}_{d}m_x+\vec{H}_{CPW})$$ \hat{x}$ +$\vec{H}_{n}$. Neglecting the effect of layers between the CPW and free layer, $\vec{H}_{CPW}$ can be approximately expressed as in\cite {dixit2012spintronic, kumar2016coherent}: $\vec{H}_{CPW}\approx I_{ac}/ 2w_{CPW}$ where $I_{ac}$ is the AC current passed through the CPW.  When driven at resonance ($w$=$w_{rn}$=$\sqrt{k_{xn}/m_{x}}$), HBAR becomes a low impedance path with $Z$($w_{rn}$)=$R_{xn}$/$\eta_{HBAR}^2$.  Therefore, the AC current becomes $I_{ac}$=($A_{amp}$$V_{STNO}$ $\eta_{HBAR}^2$)/($R_{xn}$+$R_{CPW}$) that is used to generate an AC magnetic field through the CPW where $R_{CPW}$ is assumed to be 1k$\Omega$.   Similarly, the generated magnetic field causes a phase shift and oscillation frequency changes to compensate the phase shift.  Therefore, the combination of high Q filtering and delayed feedback improves the linewidth significantly.  Moreover, this approach provides the most efficient feedback approach even though it requires additional process steps to fabricate the CPW on top of the MTJ.  It also provides an electrical isolation between the MTJ and HBAR while achieving the feedback.   
\begin{figure*}
\centering
\includegraphics[width=0.85\linewidth] {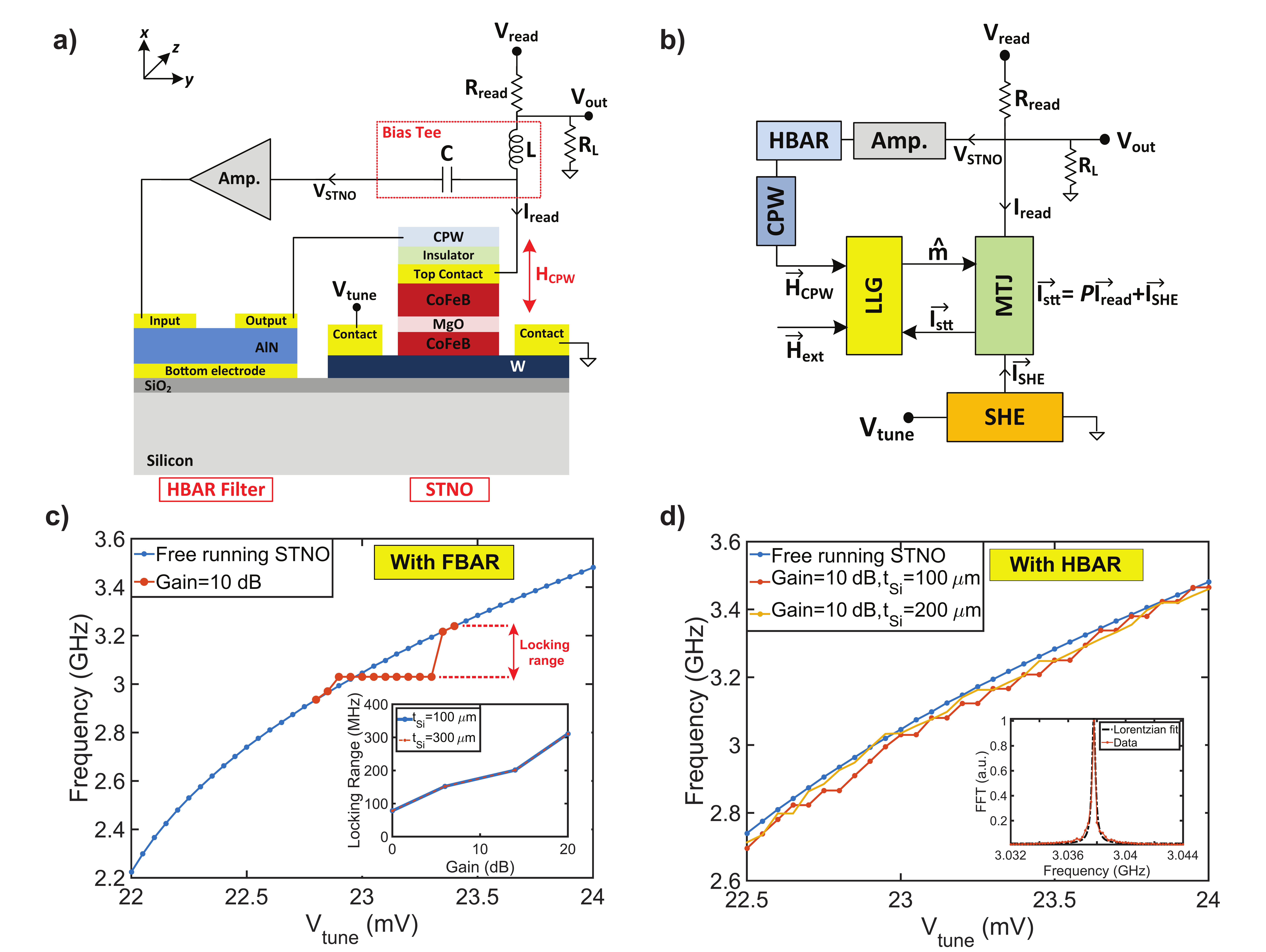}
\caption{\textbf{Implementation of MASH oscillator with magnetic field feedback:} a) A three-terminal MTJ with a CPW and a 2-port  HBAR are fabricated on the same silicon substrate, and coupled to each other with the magnetic field.   (b)  Modular modeling of transport, magnetization, and resonator dynamics. c) Locking range versus the amplifier gain for two different substrate thicknesses (100 $\mu$m and 200 $\mu$m). The thicker substrate ($t_{Si}$=200 $\mu$m) does not affect the gain needed for locking. (d) Tunability can be achieved with 22 MHz or 43 MHz steps depending on the substrate thickness using the same amplifier gain by locking to the nearest high Q peaks of the HBAR.}
\label{fig:magnetic}
\end{figure*} 

Figure 5 (c) presents  the locking range for different amplifier gains and substrate thicknesses (100 $\mu$m and 200 $\mu$m) simulated with an FBAR using the magnetic field feedback.  Gain stage is only used to compensate the transmission losses of the HBAR and even a gain of 10 dB is sufficient for the feedback, much lower compared to other feedback methods.  The locking range is independent of the substrate thickness and it is not necessary to increase the amplifier gain for thicker substrates ($t_{Si}$=200 $\mu$m) unlike the strain feedback.  Figure 5 (d) shows the tunability of  MASH oscillator with magnetic feedback where the tunability is provided with 22 MHz or 43 MHz steps depending on the substrate thickness ($t_{Si}$=100 $\mu$m or 200 $\mu$m) by locking the STNO to the nearest resonance peak of the HBAR.

\section*{Discussion}

\begin{figure*} [!t]
\includegraphics[width=0.70\linewidth] {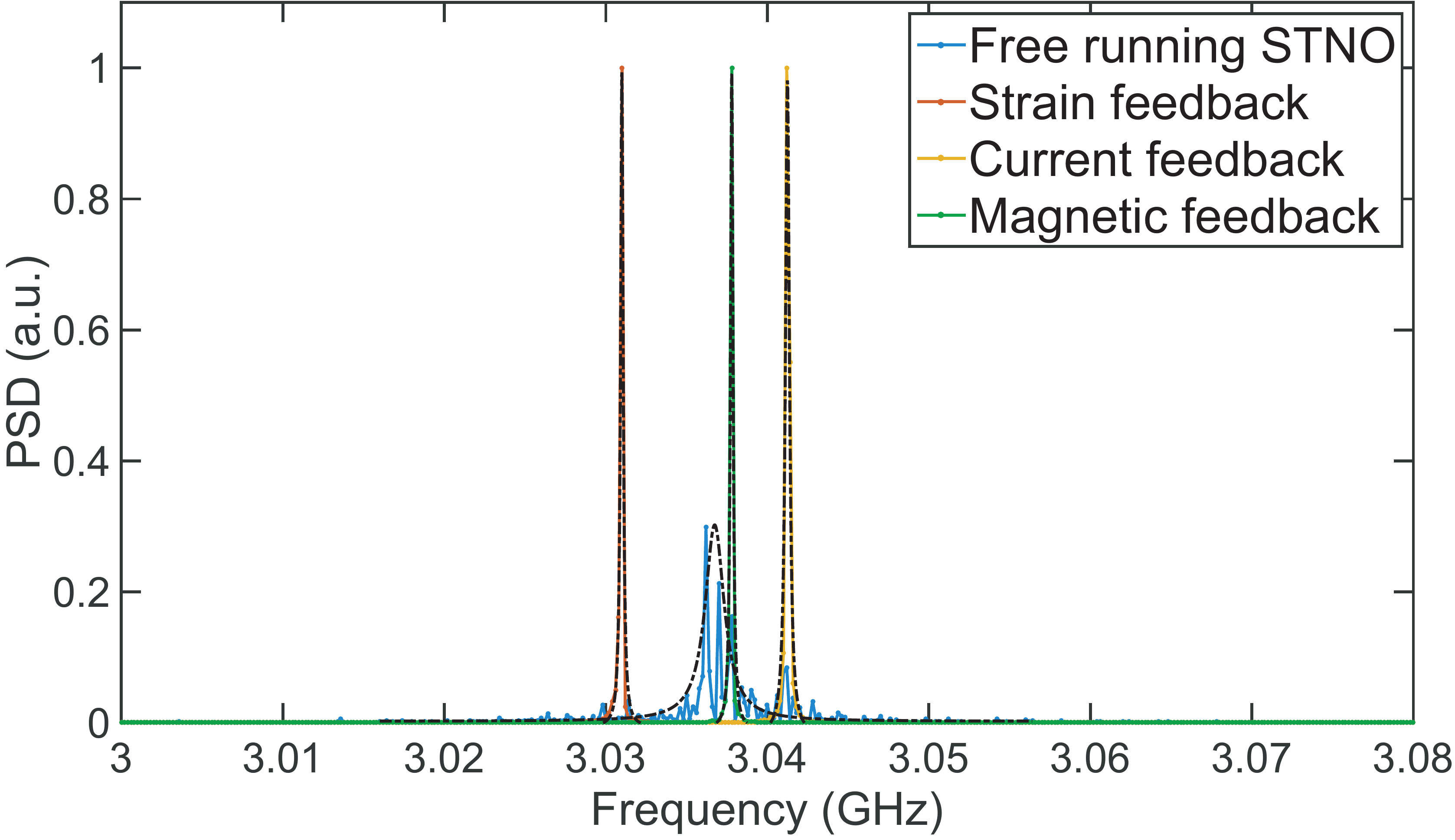} 
\caption{\textbf{Simulated PSD (Power Spectral Density) spectrums of free running STNO and MASH oscillators with different feedback methods.} Linewidth and Q of the oscillators are extracted by fitting a Lorentzian function to PSD spectrums.  MASH oscillators exhibit a 15X enhancement in the linewidth compared to the free running STNO, exceeding Q of 10000 with all feedback methods.  PSD spectrums of MASH oscillators are intentionally shifted for a clear comparison. PSD spectrum clearly shows that there are no side peaks caused by the other resonance peaks of the HBAR.}
\label{fig:FFT}
\end{figure*}  

\begin{table*} 
\centering
\resizebox{0.65\linewidth}{!}{%
\begin{tabular}{|l|l|l|l|l|l|}
\hline
Feedback &   Implementation & Gain & Linewidth & Osc. $Q$ & CMOS Compatible \\
\hline
Free running & 3-T MTJ  & N/A &  2.7 MHz & 1147 & Yes, single chip \\
\hline
Strain  & 3-T MTJ-1 port HBAR & 44 dB  & 191 kHz & 15834 & Yes, single chip  \\
\hline
Current  & 3-T MTJ-2 port HBAR & 26 dB  & 293 kHz & 10376 & Yes, single chip \\
\hline
Current  & 3-T MTJ-2 port Sapphire HBAR & 26 dB  & 55 kHz & 52825 & No, two chips  \\
\hline
Magnetic  & 3-T MTJ-2 port HBAR- CPW & 10 dB &  186 kHz & 12274& Yes, single chip  \\
\hline
\end{tabular}}
\caption{ \label{tab:Table1} Summary of results with a free running STNO and MASH oscillators implemented using different feedback methods. Linewidth is measured at 3 GHz. Gain column indicates the necessary amplitude for feedback. Osc. Q is the oscillator quality factor. 3-T designates 3-Terminal devices.}
\end{table*}

In this paper, we proposed a new class of microwave oscillators combining the tunability of a free running STNO with high Q of HBAR, integrating both reference and tunable oscillators on the same chip with CMOS circuits.  Proposed MASH oscillators are composed of a high Q HBAR reference oscillator that is locked to a three-terminal MTJ through strain, current, or magnetic field feedback.   HBAR filters voltage oscillations across the MTJ and reinjects the cleaned signal after a certain delay, causing a delayed feedback without requiring an additional source or an external delay element.  Figure 6 compares simulated FFT spectrums of a free running STNO with MASH oscillators using different feedback methods. The combination of high Q filtering and delayed feedback exhibits a 15X enhancement in the linewidth as a single chip oscillator, reaching a Q up to 15834 at 3 GHz whereas the free running STNO has a Q of 1147.  

Table 1 summarizes the results with a free running STNO and MASH oscillators, considering the different feedback methods, amplifier gain, oscillator linewidth and oscillator Q. Strain feedback provides a solution where single or multiple STNOs can be controlled with a single HBAR at the cost of larger amplifier gain whereas current and magnetic field feedback are more energy efficient and do not require any magnetostrictive materials. The CoFeB and AlN stacks used in this work do not have high magnetostriction and piezoelectric coefficients to generate a sufficient amount of magnetic field without an amplifier.  Selection of materials with higher magnetostrictive and piezoelectric coefficients such as Terfenol and PZT may generate higher magnetic field \cite {roy2012energy, khan2014voltage, gowtham2015critical}, reducing the amplifier gain necessary for the feedback.  However, the use of CoFeB and AlN layers is still a good choice since they are well optimized and CMOS compatible.

Proper choice of the acoustic substrate is essential in order to implement a high Q HBAR since Q of the HBAR depends on the acoustic losses in the piezoelectric film and substrate.  Use of a lower loss substrate such as sapphire significantly increases the Q of the HBAR up to 22,000 as experimentally observed in one of our earlier works\cite {gosavi2017magneto}, also increasing the Q of the MASH oscillator.   We have shown a 50X enhancement in the linewidth with two-chip version of the MASH oscillators using the sapphire HBAR in  \cite {gosavi2015hbar}  at the cost of being CMOS-incompatible. 

It is also crucial to maintain the tunability of the standard STNO while improving the linewidth.  The tunability has been preserved by locking the STNO to the nearest peak of the HBAR where the substrate thickness determines the tuning-steps by defining the spacing between the HBAR modes.

\begin{table}
\resizebox{0.75\linewidth}{!}{%
\begin{tabular} {|l|l|l|}
\hline
Parameters & Value\\
\hline
$V$ & 100 nm $\times$ 60 nm $\times$ 2 nm \\
\hline
$M_{s}$, $\alpha$ & 800 emu/$cm^3$, 0.01\\
\hline
$H_{an}$, $H_{d}$ & 200 Oe, 10000 Oe   \\
\hline
$R_{MTJ}$, $P_{MTJ}$ & 400 $\Omega$, 0.6 \\
\hline
$R_{read}$, $R_{L}$ & 400 $\Omega$, 50 $\Omega$ \\
\hline
$\theta _{SHE}$, $\lambda_{sf}$, $\rho_W$\cite{hao2015giant} & 0.4, 3.5 nm, 210 $\mu\Omega cm$ \\
\hline
$l_{SHE}$, $w_{SHE}$, $t_{SHE}$\cite{hao2015giant} & 60 nm, 100 nm, 9 nm\\
\hline
$B_{eff}$\cite {gowtham2016thickness} &  $-9\times10^7$ erg/$cm^3$ \\
\hline
$H_{ext}$, $V_{read}$, $V_{tune}$ & 300 Oe, 10 mV, 15-50 mV   \\
\hline
$E_{Si}$, $\rho_{Si}$, $\epsilon_{AlN}$, $e_{33}$  & 170 GPa, 2330 g/$cm^3$, 9, 1.55 C/$m^2$  \\
\hline
$r_{HBAR}$, $t_{AlN}$ & 75 $\mu$m, 1 $\mu$m \\
\hline
$w_{CPW}$, $R_{CPW}$ & 100 nm, 1 k$\Omega$ \\
\hline
$t_{sim}$, $t_{step}$ (SPICE)  & 10 $\mu$s, 0.1 ps   \\
\hline
$f_{min}$, $f_{max}$, $N$ (FFT)  & 0.1 MHz, 13.1 GHz, 262144   \\
\hline
\end{tabular}}
\caption{\label{tab:Table2} Device and material parameters used in all simulations.}
\end{table}

\section*{Methods}
MASH oscillators are implemented by combining the sLLG equation for magnetization dynamics and transport equations for SHE and MTJ self-consistently with the BVD circuit for the HBAR using the modular approach framework \cite {camsari2015modular}. HSPICE is used to solve the coupled differential equations of the oscillator network simultaneously.  All simulations are done using the transient noise analysis in HSPICE with a simulation time of $t_{sim}$= 10 $\mu$s and a step time of $t_{step}$=0.1 ps.  $t_{step}$ is selected as small as possible to completely eliminate numerical noise in the simulations.   Frequency spectrum is obtained by taking the FFT of the time domain signal. Initial 0.1 $\mu$s of the data is excluded while converting the time domain signal to the FFT. The minimum and maximum frequency of FFT are calculated by $f_{min}=(1/\Delta t)=1/(t_{stop}+t_{start})$ and $f_{max}=0.5 \times N \times f_{min}$ where $N$ is number of points.  In our simulations, we used $f_{min}$=0.1 MHz and $f_{max}$=13.1 GHz for N=262144.  PSD is plotted by having the square of absolute of FFT. Linewidth ($\Delta$f) of the oscillators is extracted by fitting a Lorentzian function to the PSD data, and Q is calculated using $f_{r}$/$\Delta$f.  Table 2 presents device and material parameters used in all simulations.

\providecommand{\latin}[1]{#1}
\providecommand*\mcitethebibliography{\thebibliography}
\csname @ifundefined\endcsname{endmcitethebibliography}
  {\let\endmcitethebibliography\endthebibliography}{}

%
\end{document}